\documentclass{PoS}

\usepackage{amsmath}

\title{Differential and total cross sections of high energy proton-proton scattering in holographic QCD}

\ShortTitle{High energy proton-proton scattering in holographic QCD}

\author{\speaker{Akira Watanabe}\\
Institute of High Energy Physics and Theoretical Physics Center for Science Facilities, Chinese Academy of Sciences, Beijing 100049, People's Republic of China\\
and\\
School of Nuclear Science and Technology, University of Chinese Academy of Sciences, Beijing 100049, People's Republic of China\\
E-mail: \email{akira@ihep.ac.cn}}

\abstract{
The analysis on the high energy proton-proton scattering is presented, focusing on the Regge regime and considering the Pomeron exchange to describe the involved nonperturbative dynamics in the framework of holographic QCD.
We combine the Reggeized spin-2 particle propagator and the proton gravitational form factor, which is obtained from the bottom-up AdS/QCD model, and calculate the differential and total cross sections.
We explicitly demonstrate the comparison between our calculations and the currently available experimental data including the recent ones measured at $\sqrt{s} = 13$~TeV by the TOTEM collaboration at the LHC.
It is shown that our results are consistent with the data, which implies that the present framework works in the considered TeV scale.
}

\FullConference{XXVII International Workshop on Deep-Inelastic Scattering and Related Subjects - DIS2019\\
		8-12 April, 2019\\
		Torino, Italy}

\begin{document}

\section{Introduction}
Understanding the quark-gluon structure of the proton is one of the most important problems in hadron physics, and the high energy scattering experiments with proton beams have played a pivotal role in deepening our understandings for decades.
The proton structure is usually described by the parton distribution functions (PDFs) which include two kinematic variables, the Bjorken variable $x$ and the probe scale $Q^2$.
Following the factorization theorem and combining the partonic hard cross sections, which are basically calculable with the perturbative technique in QCD, and PDFs, one can obtain the hadron scattering cross sections which can be compared with experimental data.
Since PDFs are nonperturbative physical quantities, the hadronic cross sections are in principle not calculable by the direct use of QCD.
Hence, effective approaches are required to theoretically make predictions for those cross sections.

In this brief report, we present based on Ref.~\cite{Xie:2019soz} our analysis on the high energy proton-proton scattering in the framework of holographic QCD, which is an effective approach constructed based on the AdS/CFT correspondence.
We consider the forward scattering and focus on the Regge regime in which the condition $s \gg |t|$, where $s$ and $t$ are the Mandelstam variables, is satisfied.
To investigate the proton structure, the deep inelastic scattering (DIS) has been intensively studied, in which the scale $Q^2$ is the four-momentum squared of the probe photon and the Bjorken $x$ is expressed as $x = Q^2/s$.
Since the typical scale in the proton-proton scattering is the proton mass $m_p \sim 1$~GeV which is much smaller than the considered $\sqrt{s}$, the gluon dynamics in the small $x$ region gives a dominant contribution to the cross sections in this study.

It is known that such the small $x$ dynamics can be well described by considering the Pomeron exchange~\cite{Donnachie:1992ny}, and this picture can be realized within the framework of holographic QCD by taking into account the Reggeized spin-2 particle which can be interpreted as the Reggeized graviton.
A lot of related studies have been done~\cite{Brower:2006ea}, and various phenomenological applications have also been attempted so far~\cite{Watanabe:2012uc,Watanabe:2013spa,Watanabe:2015mia,Watanabe:2018owy}.
In this study, we develop the preceding work done by the authors of Ref.~\cite{Domokos:2009hm} in which they investigated the elastic proton-proton scattering at high energies by applying the Reggeized spin-2 particle propagator and the dipole form factor.
To realize a more consistent description of the proton-proton scattering, we improve the treatment of the Pomeron-proton coupling, employing the proton gravitational form factor which can be obtained from the bottom-up AdS/QCD model of the nucleon~\cite{Abidin:2009hr}.
Since gravitational form factors are extracted from the corresponding energy momentum tensor matrix elements, this is a proper way to describe the Pomeron/graviton couplings.

While there are four adjustable parameters in the preceding study, our model includes only three parameters in total, which is an advantage of our approach.
This is because the previously used dipole form factor includes the dipole mass as a free parameter, but the proton gravitational form factor we use does not bring any parameter.
Those parameters are determined by a numerical fit, taking into account the experimental data of both the differential and total cross sections simultaneously.
It is explicitly shown that the currently available data at high energies, including the recent ones measured at $\sqrt{s} = 13$~TeV by the TOTEM collaboration at the LHC~\cite{Antchev:2017dia,Antchev:2017yns,Antchev:2018edk}, can be well reproduced within our model, which implies that the present framework works in the considered TeV scale.
An interesting observation which should be emphasized is that our resulting total cross section is consistent not only with the data but also with the empirical fit performed by the COMPETE collaboration~\cite{Cudell:2002xe}, although the saturation effect, which is expected to be seen in the high $s$ region, is not taken into account in the present model.

\section{Model setup}
In this study, we combine the Reggeized spin-2 particle propagator and the proton gravitational form factor, and calculate the differential and total cross sections.
Here we start with the spin-2 glueball exchange, and then introduce the expressions of the Reggeized version, in which contributions from the higher spin states are included.

The spin-2 glueball propagator with its mass $m_g$ is expressed by $d_{\alpha \beta \gamma \delta} (k) / (k^2 - m_g^2)$, and $d_{\alpha \beta \gamma \delta}$ is given by
\begin{equation}
\begin{split}
&\frac{1}{2} (\eta_{\alpha \gamma} \eta_{\beta \delta} + \eta_{\alpha \delta} %
\eta_{\beta \gamma}) - \frac{1}{2m_g^2}(k_{\alpha} k_{\delta} %
\eta_{\beta \gamma} + k_{\alpha} k_{\gamma}\eta_{\beta \delta} + k_{\beta} %
k_{\delta} \eta_{\alpha \gamma} + k_{\beta} k_{\gamma} \eta_{\alpha \delta}) \\%
& + \frac{1}{24}\left[\left( \frac{k^2}{m_g^2} \right)^2 - 3\left(\frac{k^2}{m_g^2} %
\right) - 6\right] \eta_{\alpha \beta} \eta_{\gamma \delta} %
- \frac{k^2 - 3 m_g^2}{6 m_g^4}(k_{\alpha}k_{\beta}\eta_{\gamma \delta} %
+ k_{\gamma} k_{\delta} \eta_{\alpha \beta}) %
+ \frac{2k_{\alpha}k_{\beta}k_{\gamma}k_{\delta}}{3 m_g^4}.
\end{split}
\label{eq:propagator}
\end{equation}
The proton-glueball-proton coupling is extracted from the energy-momentum tensor matrix element $\langle p', s'| T_{\mu \nu} (0) | p, s \rangle$ which is expressed with three form factors as
\begin{equation}
\bar u (p', s') \biggl[ A (t) \frac{ \gamma_\alpha P_\beta %
+ \gamma_\beta P_\alpha}{2} + B(t) \frac{i (P_\alpha \sigma_{\beta \rho} %
+ P_\beta \sigma_{\alpha \rho}) k^\rho}{4m_p} %
+ C(t) \frac{(k_\alpha k_\beta - \eta_{\alpha \beta} k^2)}{m_p} \biggr] u (p, s).
\label{eq:matrix_element}
\end{equation}
Using the above two expressions, one can calculate the amplitude.
Focusing on the Regge regime, many terms multiplied by $|t| / s$ can be neglected, and the differential cross section is obtained as
\begin{equation}
\frac{d \sigma}{dt} %
= \frac{\lambda^4 s^2 A^4 (t)}{16 \pi (t - m_g^2)^2}.
\label{eq:differential_cross_section}
\end{equation}

Since this includes only the contribution of the lightest state exchange, the propagator part needs to be Reggeized to include the higher spin state contributions.
Following the procedure explained in Ref.~\cite{Domokos:2009hm}, the expressions of the Reggeized version for the differential and total cross sections are obtained as
\begin{equation}
\frac{ d \sigma}{dt} = %
\frac{\lambda^4 A^4 (t) \Gamma^2 [ - \chi ] \Gamma^2 \left[ 1 - \frac{ \alpha_c (t)}{2} \right] } {16 \pi \Gamma^2 \left[ \frac{\alpha_c (t)}{2} - 1 - \chi \right] } %
\left( \frac{\alpha'_c s}{2}\right)^{2 \alpha_c (t) - 2},
\label{eq:differential_cross_section_Reggeized}
\end{equation}
\begin{equation}
\sigma_{tot} = %
\frac{\pi \lambda^2 \Gamma [ - \chi ]}{\Gamma \left[ \frac{\alpha_c (0)}{2} \right] \Gamma \left[ \frac{\alpha_c (0)}{2} - 1 - \chi \right] } %
\left( \frac{\alpha_c' s}{2} \right)^{\alpha_c (0) - 1},
\label{eq:total_cross_section_Reggeized}
\end{equation}
respectively, where $\alpha_{c} (x) = \alpha_{c} (0) + \alpha'_{c} x$ and $\chi = \alpha_{c} (s) + \alpha_{c} (t)  + \alpha_{c} (u) = 4 \alpha'_{c} m^2 + 3 \alpha_{c} (0)$.
There are three adjustable parameters, $\lambda$, $\alpha_{c} (0)$, and $\alpha'_{c}$, which are to be determined with the experimental data.

In our model, the proton-Pomeron-proton coupling is specified by the proton gravitational form factor, $A (t)$ in Eq.~\eqref{eq:differential_cross_section_Reggeized} in this study.
This form factor can be calculated using the bottom-up AdS/QCD model~\cite{Abidin:2009hr}, and its applications to the analysis on DIS~\cite{Watanabe:2012uc,Watanabe:2013spa} and on the proton-proton total cross section~\cite{Watanabe:2018owy} have been done so far.
In the bottom-up AdS/QCD model, one needs to set a cutoff in the infrared region of the AdS space to introduce the QCD scale, and there are two ways to realize it.
In the hard-wall model, a sharp cutoff is imposed, and the AdS geometry is smoothly cut off by utilizing the dilaton field in the soft-wall model.
In this study, we consider the both cases for the form factor to obtain the numerical results.

\section{Numerical results}
Taking into account the currently available experimental data for the differential and total cross sections, the three parameters in the model are determined by a numerical fit.
Since we consider only the Pomeron exchange contribution in this study and the diffractive minimum around $|t| = 0.47$~GeV$^2$ has been observed recently in the differential cross section measurement by the TOTEM collaboration~\cite{Antchev:2018edk}, we select the data in the kinematic regime, where $\sqrt{s} \ge 546 $~GeV and $|t| < 0.45$~GeV$^2$.
As the results of the fit, the $\chi^2 / d.o.f.$ values, 1.317 and 1.355 for the soft-wall and hard-wall model, respectively, were obtained.
We show only the soft-wall model results hereafter in this report.

\begin{figure}[tb]
\centering
\includegraphics[width=0.99\textwidth]{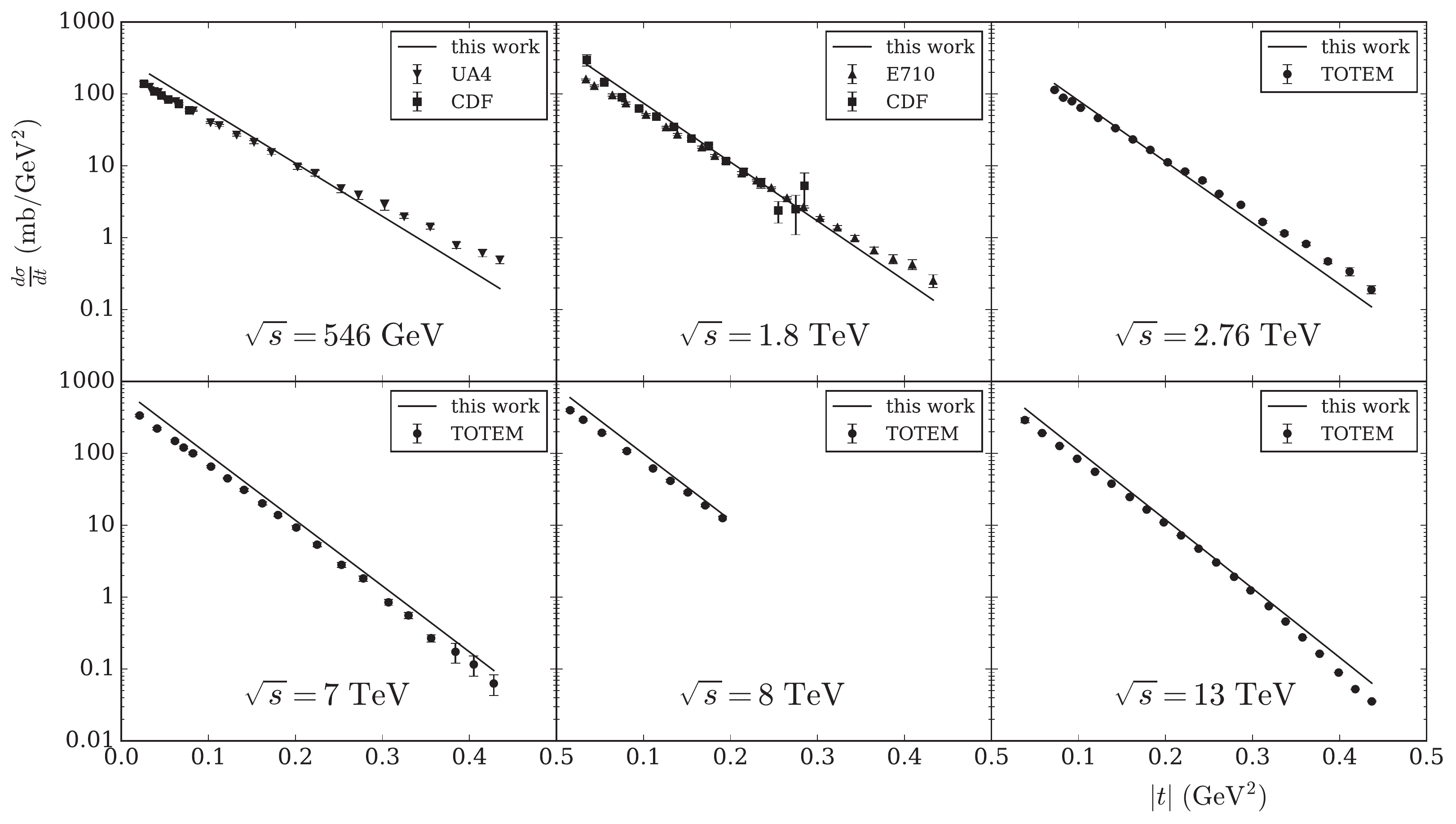}
\caption{
The differential cross section as a function of $|t|$ for the various values of $\sqrt{s}$.
The solid lines represent our calculations, and the experimental data measured by the several collaborations are depicted with their errors.
}
\label{fig:dc}
\end{figure}

\begin{figure}[tb]
\centering
\includegraphics[width=0.67\textwidth]{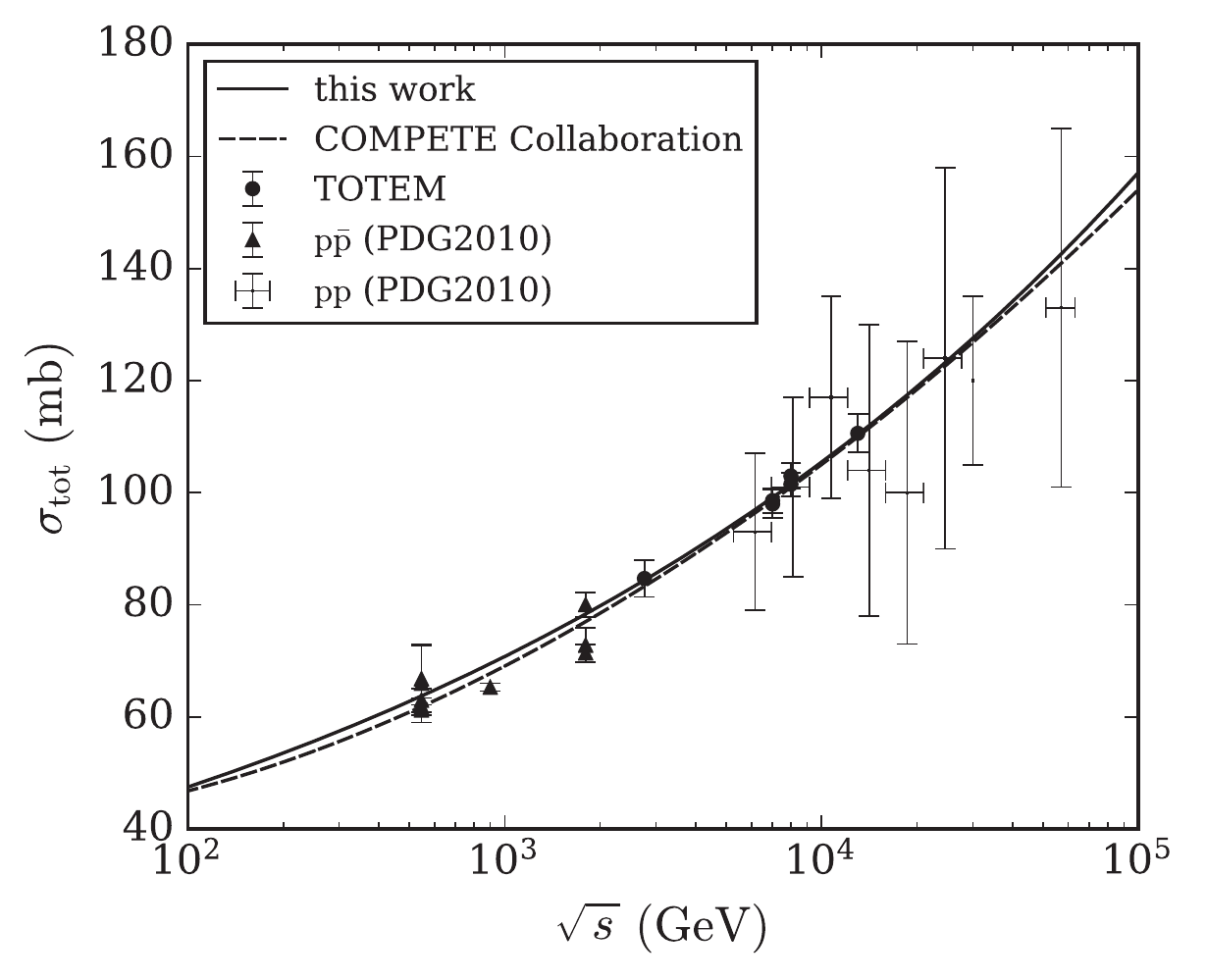}
\caption{
The total cross section as a function of $\sqrt{s}$.
The solid and dashed curves denote our calculation and the empirical fit obtained by the COMPETE collaboration~\cite{Cudell:2002xe}, respectively.
The experimental data are depicted with their errors.
The pp data taken from PDG2010, which were extracted from cosmic-ray experiments, were not used for the parameter fit.
}
\label{fig:tcs}
\end{figure}

We display in Fig.~\ref{fig:dc}
our calculations for the differential cross section.
It is seen that our results are consistent with the data measured by the various collaborations.
Then, we show in Fig.~\ref{fig:tcs}
the resulting total cross section, compared with the data and the empirical fit obtained by the COMPETE collaboration~\cite{Cudell:2002xe}.
We plot several pp data collected by the Particle Data Group (PDG) in 2010, which were extracted from cosmic-ray experiments, together with the other collider experiment data, but they were not used for the parameter fit because of their huge uncertainties.
One can see from the figure that our calculation is in agreement with both the data and the empirical fit in the whole considered kinematic region.

\section{Summary}
We have studied the high energy proton-proton scattering, focusing on the Regge regime and considering the Pomeron exchange, in the framework of holographic QCD.
Combining the Reggeized spin-2 particle propagator and the proton gravitational form factor obtained from the bottom-up AdS/QCD model, we calculate the differential and total cross sections.
In our model setup, there are only three adjustable parameters because this form factor does not bring any free parameter, which shows an advantage over the preceding work.

Our calculations for both the differential and total cross sections are consistent with the currently available experimental data measured by various collaborations, including the recent TOTEM ones at $\sqrt{s} = 13$~TeV.
It should be noted that our resulting total cross section is also in good agreement with the empirical fit obtained by the COMPETE collaboration.
In the high energy region, which may be in the TeV scale, the saturation effect is expected to be seen in experimental data.
For instance, it is expected that one can observe a suppression due to the effect in the proton-proton total cross section data.
Since this effect is not taken into account in the present model, more data at higher energies would help to make constraints on phenomenological models and also to deepen our understandings about the gluon recombination in high energy QCD.

The results we obtained in this study imply that the present model works well in the considered TeV scale.
However, there are still missing contributions, for instance, those from the Reggeon exchange, the multi-Pomeron exchange, the Odderon exchange, and so on, besides the saturation effect.
Further improvements of the model itself are certainly needed.
Moreover, further applications to other high energy scattering processes would also be necessary to investigate the reliability of the model and its applicable limits.

\acknowledgments{
The work of the author was supported by Chinese Academy of Sciences (CAS) President's International Fellowship Initiative under Grant No. 2019PM0124 and partially by China Postdoctoral Science Foundation under Grant No. 2018M641473.
}

\bibliographystyle{JHEP}
\bibliography{ref}

\providecommand{\href}[2]{#2}\begingroup\raggedright\begin{thebibliography}{10}

\bibitem{Xie:2019soz}
W.~Xie, A.~Watanabe and M.~Huang, \emph{{Elastic proton-proton scattering at
  LHC energies in holographic QCD}},
  \href{https://arxiv.org/abs/1901.09564}{{\ttfamily 1901.09564}}.

\bibitem{Donnachie:1992ny}
A.~Donnachie and P.~V. Landshoff, \emph{{Total cross-sections}},
  \href{https://doi.org/10.1016/0370-2693(92)90832-O}{\emph{Phys. Lett.}
  {\bfseries B296} (1992) 227}
  [\href{https://arxiv.org/abs/hep-ph/9209205}{{\ttfamily hep-ph/9209205}}].

\bibitem{Brower:2006ea}
R.~C. Brower, J.~Polchinski, M.~J. Strassler and C.-I. Tan, \emph{{The Pomeron
  and gauge/string duality}},
  \href{https://doi.org/10.1088/1126-6708/2007/12/005}{\emph{JHEP} {\bfseries
  12} (2007) 005} [\href{https://arxiv.org/abs/hep-th/0603115}{{\ttfamily
  hep-th/0603115}}].

\bibitem{Watanabe:2012uc}
A.~Watanabe and K.~Suzuki, \emph{{Transition from soft- to hard-Pomeron in the
  structure functions of hadrons at small-$x$ from holography}},
  \href{https://doi.org/10.1103/PhysRevD.86.035011}{\emph{Phys. Rev.}
  {\bfseries D86} (2012) 035011}
  [\href{https://arxiv.org/abs/1206.0910}{{\ttfamily 1206.0910}}].

\bibitem{Watanabe:2013spa}
A.~Watanabe and K.~Suzuki, \emph{{Nucleon structure functions at small $x$ via
  the Pomeron exchange in AdS space with a soft infrared wall}},
  \href{https://doi.org/10.1103/PhysRevD.89.115015}{\emph{Phys. Rev.}
  {\bfseries D89} (2014) 115015}
  [\href{https://arxiv.org/abs/1312.7114}{{\ttfamily 1312.7114}}].

\bibitem{Watanabe:2015mia}
A.~Watanabe and H.-n. Li, \emph{{Photon structure functions at small $x$ in
  holographic QCD}},
  \href{https://doi.org/10.1016/j.physletb.2015.10.069}{\emph{Phys. Lett.}
  {\bfseries B751} (2015) 321}
  [\href{https://arxiv.org/abs/1502.03894}{{\ttfamily 1502.03894}}].

\bibitem{Watanabe:2018owy}
A.~Watanabe and M.~Huang, \emph{{Total hadronic cross sections at high energies
  in holographic QCD}},
  \href{https://doi.org/10.1016/j.physletb.2018.11.042}{\emph{Phys. Lett.}
  {\bfseries B788} (2019) 256}
  [\href{https://arxiv.org/abs/1809.02515}{{\ttfamily 1809.02515}}].

\bibitem{Domokos:2009hm}
S.~K. Domokos, J.~A. Harvey and N.~Mann, \emph{{The Pomeron contribution to p p
  and p anti-p scattering in AdS/QCD}},
  \href{https://doi.org/10.1103/PhysRevD.80.126015}{\emph{Phys. Rev.}
  {\bfseries D80} (2009) 126015}
  [\href{https://arxiv.org/abs/0907.1084}{{\ttfamily 0907.1084}}].

\bibitem{Abidin:2009hr}
Z.~Abidin and C.~E. Carlson, \emph{{Nucleon electromagnetic and gravitational
  form factors from holography}},
  \href{https://doi.org/10.1103/PhysRevD.79.115003}{\emph{Phys. Rev.}
  {\bfseries D79} (2009) 115003}
  [\href{https://arxiv.org/abs/0903.4818}{{\ttfamily 0903.4818}}].

\bibitem{Antchev:2017dia}
{\scshape TOTEM} collaboration, G.~Antchev et~al., \emph{{First measurement of
  elastic, inelastic and total cross-section at $\sqrt{s}=13$ TeV by TOTEM and
  overview of cross-section data at LHC energies}},
  \href{https://doi.org/10.1140/epjc/s10052-019-6567-0}{\emph{Eur. Phys. J.}
  {\bfseries C79} (2019) 103}
  [\href{https://arxiv.org/abs/1712.06153}{{\ttfamily 1712.06153}}].

\bibitem{Antchev:2017yns}
{\scshape TOTEM} collaboration, G.~Antchev et~al., \emph{{First determination
  of the $\rho$ parameter at $\sqrt{s}=13$ TeV -- probing the existence of a
  colourless three-gluon bound state}},
  \href{https://arxiv.org/abs/1812.04732}{{\ttfamily 1812.04732}}.

\bibitem{Antchev:2018edk}
{\scshape TOTEM} collaboration, G.~Antchev et~al., \emph{{Elastic differential
  cross-section measurement at $\sqrt{s}=13$ TeV by TOTEM}},
  \href{https://arxiv.org/abs/1812.08283}{{\ttfamily 1812.08283}}.

\bibitem{Cudell:2002xe}
{\scshape COMPETE} collaboration, J.~R. Cudell, V.~V. Ezhela, P.~Gauron,
  K.~Kang, {\relax Yu}.~V. Kuyanov, S.~B. Lugovsky et~al., \emph{{Benchmarks
  for the forward observables at RHIC, the Tevatron Run II and the LHC}},
  \href{https://doi.org/10.1103/PhysRevLett.89.201801}{\emph{Phys. Rev. Lett.}
  {\bfseries 89} (2002) 201801}
  [\href{https://arxiv.org/abs/hep-ph/0206172}{{\ttfamily hep-ph/0206172}}].

\end{thebibliography}\endgroup



\end{document}